\documentclass[prl,twocolumn,showpacs,floatfix,superscriptaddress]{revtex4}
\usepackage[T1]{fontenc}
\usepackage[cp1250]{inputenc}
\usepackage{psfrag}
\usepackage{graphicx}
\usepackage[english]{babel}
\newcommand{\bq}{\begin{equation}}
\newcommand{\eq}{\end{equation}}
\newcommand{\ba}{\begin{eqnarray}}
\newcommand{\ea}{\end{eqnarray}}
\begin{document}
\title{Tuning the shape of the condensate in spontaneous symmetry breaking}
\author{B. Waclaw}
\affiliation{Institut f\"ur Theoretische Physik, Universit\"at Leipzig, Postfach 100\,920, 04009 Leipzig, Germany }
\author{J. Sopik}
\affiliation{School of Engineering and Science, Jacobs
University, P.O. Box 750561, 28725 Bremen, Germany}
\author{W. Janke$^1$}
\affiliation{}
\author{H. Meyer-Ortmanns$^2$}
\affiliation{}
\noindent
\begin{abstract}
We investigate what determines the shape of a particle condensate in situations when it emerges as a result of spontaneous breaking of translational symmetry.
We consider a model with particles hopping between sites of a one-dimensional grid and interacting if they are at the same or at neighboring nodes.
We predict the envelope of the condensate and the scaling of its width with the system size for various interaction potentials and show how to tune the shape from a delta-peak to a rectangular or a parabolic-like form.
\end{abstract}
\maketitle

The phenomenon of condensation is widely spread in nature and appears in equilibrium systems as well as in systems driven far from equilibrium.
Examples of equilibrium condensation are Bose-Einstein condensation or certain phase transitions in quantum gravity \cite{grav}. 
In non-equilibrium systems, the condensation manifests itself in microscopic (intracellular) and macroscopic (highway) traffic \cite{chowdhury} where it corresponds to jamming, granular flow \cite{liu}
and granular clustering \cite{majumdar}, or gelation in networks \cite{redner} where a single node takes a finite fraction of all links. 

Many of these systems can be modeled as a set of particles occupying discrete levels, or ``boxes'' on a one-dimensional grid.
The balls-in-boxes model (B-in-B) \cite{bbj}, or its non-equilibrium version, the zero-range process (ZRP) \cite{evans2000}, are well-known examples. In these models, the condensation transition happens along with a spontaneous breaking of translational symmetry.
Although these models are more abstract than realistic ones, they serve as paradigms since the stationary state can be derived analytically
as fully factorizing over the sites of the grid, or, more generally, over the nodes of an arbitrary graph \cite{2006}.
The factorization is due to ultra-local (``zero-range'') rules assumed to govern the dynamics of particles.
This makes the phase structure of the system accessible and enables comparison with experiments on condensation phenomena, at least on a qualitative level. 

As a natural consequence of the lack of interactions between particles at neighboring sites, the condensates in the ZRP or B-in-B models  always occupy a single site. 
The question then arises how the shape of the condensate changes in the presence of interactions between sites that tend to flatten out the condensate's profile but still preserve translational symmetry and conserve the current. 
In this paper we study this problem in a model similar to the one proposed
in Ref.~\cite{evans1}. 
It is related to a solid-on-solid (SOS) model \cite{sos,sos2} supplied with dynamical rules that drive the system out-of-equilibrium. Here, the steady state factorizes over {\em pairs\/} of sites which allows for nearest-neighbor interactions while making the system analytically solvable.
We shall show that although for a broad class of hopping rates the condensate becomes extended to $W\sim\sqrt{N}$ sites, where $N$ is the total number of sites, it is possible to obtain any scaling $W\sim N^\alpha$ with $0\leq\alpha\leq 1/2$. We shall also predict the shape of the condensate for some special cases. Possible applications to surface science will be mentioned in the conclusions.

We consider a ring with $N$ sites and $M$ particles of unit mass placed randomly on the sites.
Each site $i$ can carry an arbitrary number $m_i=0,\dots,M$ of particles. The dynamics is divided into two steps. As a first step, a particle may leave a randomly chosen site with probability $u(m_i|m_{i-1},m_{i+1})$ which depends on the state of neighboring sites. As a second step, the particle chooses a target to the right site with probability $r$, or to the left one with probability $1-r$.
One can show \cite{bw1} that when the hopping rate has the form
\bq
	u(m_i|m_{i-1},m_{i+1}) = \frac{g(m_i-1,m_{i-1})}{g(m_i,m_{i-1})}\frac{g(m_i-1,m_{i+1})}{g(m_i,m_{i+1})},
\eq
with $g(m,n)$ being a symmetric non-negative but otherwise arbitrary function, the steady state factorizes over pairs of sites according to
\bq
	P(m_1,\dots,m_N) = \prod_{i=1}^N g(m_i,m_{i+1}) \delta\left(\sum_{i=1}^{N}\;m_i-M\right),
	\label{eq1}
\eq
where the ring geometry implies $m_{N+1}\equiv m_1$.
The difference as compared to the model of Ref.~\cite{evans1} is that there $r=1$ and no assumption on the symmetry of $g(m,n)$ was made.
The parameter $r$ changes only the net current of particles. When $r=1/2$, the current is zero and the system is in equilibrium. Since the steady state (\ref{eq1}) does not depend on $r$, its static properties can be calculated by formally treating the system as if it was in equilibrium with the probability of a microstate given by Eq.~(\ref{eq1}).

The criterion for condensation was analyzed in Ref.~\cite{evans1} by means of the grand-canonical partition function
\bq
	Z_N(z) = \sum_{\{m_i\}} z^{\sum_i m_i} \prod_i g(m_i,m_{i+1}) = \mbox{Tr} \, T^N,
	\label{zgrand}
\eq
where $T_{mn}\equiv z^{(m+n)/2}g(m,n)$ and $z$ is the fugacity. $Z_N(z)$ must have a finite range of convergence $z_c$.
The critical value of the density $\rho = M/N$ then follows from
\bq
	\rho_c=\lim_{N\to\infty,z\to z_c^-} \frac{z}{N}\frac{\partial\ln Z_N(z)}{\partial z} = \frac{\sum_m m\phi_m^2 }{\sum_m\phi_m^2}\;,
	\label{eqh1}
\eq
where $\phi$ is an eigenvector of $T_{mn}$ to the largest eigenvalue $\lambda_{\rm max}$,
for $z=z_c$. At the critical point, $Z_N \cong \lambda_{\rm max}^N$ for large $N$.
In what follows we assume that $g(m,n)$ has been rescaled so that $z_c=1$, hence
the critical density can be obtained by diagonalizing $T_{mn}=g(m,n)$.

To investigate how the introduction of site-site interactions influences the properties of condensation, we stick to the following choice:
\bq
	g(m,n) = K(|m-n|) \sqrt{p(m)p(n)}. \label{kmnp}
\eq
When $K(x)=1$, $g(m,n)$ factorizes and we recover the ZRP. Here we assume that both $K(x)$ and $p(m)$ are some positive, decaying functions of $x$ and $m$, respectively. The choice (\ref{kmnp}) is motivated by studies on the SOS model in the context of surface roughening \cite{sos,sos2} and corresponds to the energy $E=-\ln K(|m-n|) - 1/2(\ln p(m)+\ln p(m))$ of the interface in a 1+1 dimensional surface (where the surface refers to the envelope of occupation numbers).
In Ref.~\cite{evans1}, the following choice was proposed:
\bq
  K(x)=e^{-Jx}, \;\;\;\; p(m)=e^{ U\delta_{m0}}, \label{wevans}
\eq
with parameters $J$ and $U$ generating an effective surface stiffness and a pinning potential, respectively.
For the weights (\ref{wevans}), we obtain
\bq
	\rho_c=\frac{e^{J_0}-1}{(e^{J_0}-e^{-2(J-J_0)})(e^{2(J-J_0)}-1)},
	\label{eqh4}
\eq
with $J_0=U-\ln(e^{U}-1)$.
In the limit $J \rightarrow J_0$, this agrees with the asymptotic result in Ref.~\cite{evans1}.
For $J=U=1$, Eq.~(\ref{eqh4}) gives $\rho_c\approx 0.2397$, which we also confirmed numerically.
If $J<J_0$, the critical density is infinite.

{\em The generic case} ---
For the weight functions (\ref{wevans}),
the width $W$ was estimated in Ref.~\cite{evans1} by a random-walk argument to be proportional to $\sqrt{N}$.
Here we use a different approach which allows to calculate not only $W$ but also the envelope of the condensate in the case when $K(x)$ decays exponentially or faster in $x$, and $p(m)\to 1$ for $m\to\infty$. First of all, one cannot use the partition function (\ref{zgrand}), because it is not defined in the condensed state. We assume, however,
that the system can be split into the condensate which extends over $W$ sites occupied by $M'=M-N\rho_c$ particles on the average and where average occupation numbers $\langle m_i \rangle$ grow with $N$, and a uniform background with
$\left<m_i\right> = \rho_c$. Since fluctuations in the background are finite and have no long-range correlations, the mass $M'$ cannot fluctuate more than $\sim\sqrt{N}$ so that we treat it as constant.
We can therefore assume that the probability of having the condensate extended to $W$ sites factorizes:
\ba
	P(W) &\approx & Z_{\rm b}(W) Z_{\rm c}(W) \nonumber \\
			 &\propto &\exp\left(-W\ln \lambda_{\rm max} +\ln Z_{\rm c}(W)\right), \label{znm}
\ea
where $Z_{\rm b}(W)=\lambda_{\rm max}^{N-W}$ is the partition function (\ref{zgrand}) for the background at the critical point $\rho=\rho_c$ and $Z_{\rm c}(W)$ is the partition function of the condensate extended over $W$ sites and having exactly $M'$ particles. The average extension $W$ can be determined by the maximum of Eq.~(\ref{znm}). Because $p(m)\to 1$ for large $m$, we can neglect the contribution from $p(m)$ to $Z_{\rm c}(W)$. 
We have
\bq
	Z_{\rm c}(W) \cong \sum_{\{m_k\}} \prod_{k=1}^{W+1} K(|m_k-m_{k-1}|)\delta\left(\sum_{k=1}^W m_k - M' \right), \label{z1}
\eq
where we assumed that $m_0=m_{W+1}=0$, because occupation numbers at the boundaries are small.
For large systems, the conservation law given by the delta function forces the majority of occupation numbers $m_k$ to be much larger than zero. We can thus extend the summation to negative $m_k$ and change the summation over $\{m_k\}$ into a summation over $\{d_k\}$, where $d_k=m_k-m_{k-1}$.
Defining the generating function
\ba
	G(W,\vec{u}) &=& \sum_{d_1=-\infty}^\infty \cdots \sum_{d_{W}=-\infty}^\infty \prod_{k=1}^{W} K(|d_k|) e^{d_k u_k}\nonumber \\
	&\times& \delta\left(-\sum_{k=1}^{W} kd_k - M' \right) \delta\left(\sum_{k=1}^{W} d_k \right) \label{gen}
\ea
with auxiliary fields $\vec{u}=\{u_1,\dots,u_{W}\}$,
we can rewrite Eq.~(\ref{z1}) as $Z_{\rm c}(W) \cong G(W+1,\vec{0})$. The first delta function in Eq.~(\ref{gen}) implements the mass conservation in the condensate, the second delta reflects fixed-boundary conditions. Using integral representations of both deltas, the generating function (\ref{gen}) can be evaluated in the saddle point as $G(W,\vec{u})\sim e^{F(z,v,\vec{u})}$, where
\bq
	F(z,v,\vec{u}) = -M' z + \sum_{k=1}^W \ln \tilde{K}(u_k+v-k z). \label{fzvu}
\eq
The variables $z=z(\vec{u}),v=v(\vec{u})$ are determined from the saddle-point equation $\partial_z F(z,v,\vec{u}) = \partial_v F(z,v,\vec{u})=0$, and the function $\tilde{K}(x)$ is defined by
\bq
	\tilde{K}(x) = \sum_{d=-\infty}^\infty K(|d|) e^{dx}.
	\label{ktilde}
\eq
We assume that this sum exists,
as it does for the case (\ref{wevans}).
{}From the definition of $G(W,\vec{u})$, $\left< d_k\right> = \frac{d}{du_k}F(z,v,\vec{u})|_{\vec{u}=0}$.
The symmetry of the averaged peak implies that $\left< d_k\right> = -\left< d_{W-k}\right>$ and hence $z = (2/W) v$. We obtain
\bq
	\left<d_k\right> = \left.\tilde{K}'(x)/\tilde{K}(x)\right|_{x=v(1-2k/W)}.
\eq
In addition, from the conservation law $\sum_k k \left< d_k\right> = -M'$ we have for large systems
\bq
		\frac{1}{2v^2} \int_0^{v}  \frac{x\tilde{K}'(x)}{\tilde{K}(x)} {\rm d} x= \frac{1}{w^2}, \label{ftox}
\eq
where we defined the reduced extension $w\equiv W/\sqrt{M'}$.
Now, since $\ln Z_{\rm c}(W) \cong F(2v/(w\sqrt{M'}),v,\vec{0})$, we can rewrite Eq.~(\ref{znm}) as a function of $v$:
\bq
	\frac{\ln P(W(v))}{\sqrt{M'}} = -w\ln\lambda_{\rm max} -\frac{2v}{w}  + \frac{w}{v} \int_0^{v} \ln \tilde{K}(x) {\rm d}x,
\eq
having in mind that $w$ is not an independent variable but is bound to $v$ through Eq.~(\ref{ftox}). Taking the derivative with respect to $v$ we find after some calculations that $P(W)$ has a maximum for $v_0=\tilde{K}^{-1}(\lambda_{\rm max})$. The corresponding width $W=w_0\sqrt{M'}$ of the condensate can be determined from Eq.~(\ref{ftox}) by calculating $w_0$ numerically.
Since $w_0$ is independent of the system size, we have thus shown that $W\sim \sqrt{M'} \sim \sqrt{N}$.
A detailed discussion will be presented elsewhere \cite{bw1}.

\begin{figure}
	\includegraphics[width=5.5cm]{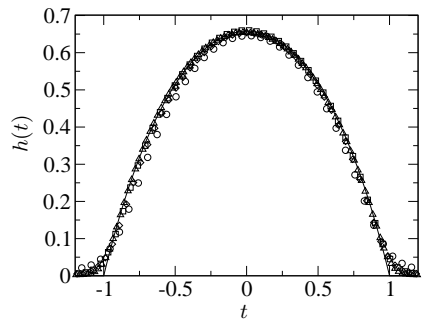}
\caption{\label{fig:shape}Comparison between the envelope of the condensate from Eq.~(\ref{htfinal}) and rescaled MC data for the weights (\ref{wevans}) with $J=U=1$ for $N=1000$, $\rho=M/N=1$ and $3$ (circles, squares) and $N=4000$, $\rho=1,3$ (diamonds, triangles).
}
\end{figure}

To calculate the envelope of the condensate,
it is convenient to define $h(t)\equiv \left<m_n\right>/\sqrt{M^\prime}$ in a rescaled variable $t=\frac{2n}{w_0\sqrt{M'}}-1$, which removes the dependence on the system size. In Fig.~\ref{fig:shape} we show that the Monte Carlo (MC) data points, obtained for different sizes, collapse onto a single curve that looks like a parabola.
To calculate it analytically, we note that $\left<m_n\right>=\sum_{k=1}^{n}\left<d_k\right>$.
Changing the summation into integration, we obtain
\bq
	h(t) = \frac{w_0}{2v_0} \ln \frac{\tilde{K}(v_0)}{\tilde{K}(v_0 t)}. \label{mnfinal}
\eq
For the special case of Eq.~(\ref{wevans}), one obtains
\bq	
h(t) = \frac{w_0}{2v_0} \ln \left( \frac{\cosh J - \cosh v_0 t}{\cosh J - \cosh v_0}\right),
	\label{htfinal}
\eq
with $w_0$, $v_0$ being functions of the parameters $J,U$. The formula for $v_0$ is particularly simple, $v_0=J-J_0$.
The comparison between Eq.~(\ref{htfinal}) for $J=U=1$, which gives
$v_0=0.5413$ and $w_0=2.2005$ (from Eq.~(\ref{ftox})), and MC simulations in Fig.~\ref{fig:shape} shows a very good agreement.

\begin{figure}
	\includegraphics[width=7cm]{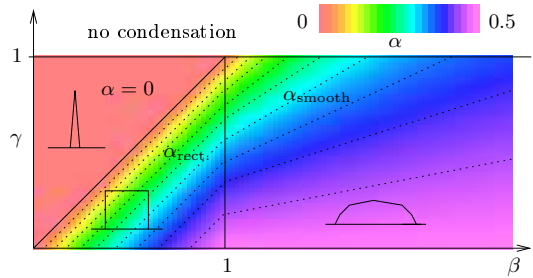}
	\caption{\label{phase1}(color online) Phase diagram for $K(x)\sim e^{-x^{\beta}}$ and $p(m)\sim e^{-m^\gamma}$. The exponent $\alpha_{\rm rect}=(\beta-\gamma)/(\beta-\gamma+1)$ for rectangular and $\alpha_{\rm smooth}=(\beta-\gamma)/(2\beta-\gamma)$ for smooth condensates.
	The dotted lines show $\alpha=0.05,0.1,\dots,0.45$ (from left to right).
	}
\end{figure}

{\em Extension different from $\sqrt{N}$} ---
So far we considered the situation which, in the language of the SOS model, corresponds to interactions between neighboring sites which decay rapidly with the difference of heights, and the pinning potential $\ln p(m)$ localized near zero.
In this case, the extension scales generically as $\sim\sqrt{N}$. We will show now that the exponent $\alpha$ in the extension $W\sim N^\alpha$ can be tuned to $\alpha\neq 1/2$ for the following choice of the weight functions: $K(x)\sim e^{-x^\beta}$ and $p(m)\sim e^{-m^\gamma}$ with $\beta, \gamma >0$. Such weights translate to the SOS energy $E=\sum_i |m_i-m_{i-1}|^\beta + m_i^\gamma$. The exponent $\gamma=1$ corresponds to a constant, e.g., a gravitational field \cite{sos2} acting on the envelope, while $\gamma<1$ and $\gamma>1$ corresponds to some attractive force, decreasing or increasing with $m$, respectively.
Although the series (\ref{ktilde}) is convergent, $\lim_{m\to\infty} p(m)\to 0$, so that we cannot use the previous method. However, we can proceed as follows.
{}From the condensation criterion we obtain that for $\gamma>1$ there is no condensation. The condensate emerges above some $\rho_c$ only for $\gamma<1$, the value of $\beta$ does not matter.
If the condensate was localized at a single site, its statistical weight would be $	P_1 \sim N c^{N-1} p(N) K^2(N)$.
Here $c$ stands for the contribution of background sites. For a two-site condensate, both sites will be almost equally occupied because every difference $\epsilon$ is suppressed as $K(\epsilon)$, and we have
$P_2 \sim N c^{N-2} K^2\left(\frac{N}{2}\right) p^2\left(\frac{N}{2}\right)$.
We thus obtain $\ln P_1/P_2 \approx - c_\beta N^\beta + c_\gamma N^\gamma$ with some constants $c_\beta,c_\gamma>0$ for $\gamma<1$. This means that for $\gamma>\beta$, the ratio $P_1/P_2 \to \infty$ as $N\to\infty$ and the condensate is localized. However, for $\gamma<\beta$, the contribution from the two-site condensate is larger. Similarly, one can show that the contribution $P_n$ for an $n$-site condensate grows with $n$, so the condensate must be extended.
Denoting by $P(W)$ the probability for an extension $W$, we can estimate $\ln P(W)$
as follows, dropping inessential constants.
First, there is a negative $-W$ term that accounts for the $W$ excluded background sites (cf.\ Eq.~(\ref{znm})).
An analogous term comes from fluctuations (we assume that the entropy is additive).
Second, every site in the condensate contributes as $p(N/W)$ (because $M'\sim N$), so there is a $-W(N/W)^\gamma$ term.
The third term comes from the weight factor $K(x)$ and depends on the condensate shape. As originally suggested by numerical experiments, we only have to distinguish two types (see below).
For shapes with a {\em smooth\/} envelope $h(t)$, the contribution is
\bq
\sim W \int \ln K(2Nh'(t)/W^2) {\rm d}t \sim - W\left(N/W^2\right)^\beta,
\eq
while for a {\em rectangular\/} shape, where $h(t)$ has a discontinuity at the borders, the contribution is
\bq
\sim 2\ln K(N/W) \sim -\left(N/W\right)^\beta.
\eq
Taking all terms together,
the extension $W$ follows by searching the maximum of the larger of the two values $P(W)$:
\ba
	\ln P_{\rm smooth} \sim -W - W(N/W)^\gamma- W\left(N/W^2\right)^\beta, \label{smooth-shape} \\
	\ln P_{\rm rectangular} \sim -W - W(N/W)^\gamma -\left(N/W\right)^\beta. \label{rectangular-shape}
\ea
Assuming $W\sim N^\alpha$, this gives $\alpha_{\rm rect}=(\beta-\gamma)/(\beta-\gamma+1)$ for the rectangular, and $\alpha_{\rm smooth}=(\beta-\gamma)/(2\beta-\gamma)$ for the smooth condensate, respectively, a non-trivial exponent that reminds of finite-size scaling at second-order phase transitions.
Mixed shapes, partially smooth but with some discontinuities,
would give a  contribution
which is always smaller than the larger value of Eqs.~(\ref{smooth-shape}) and (\ref{rectangular-shape}).
Thus we can have only these two types of extended shapes, and a localized one,
depending on the values of $\beta$ and $\gamma$.
In Fig.~\ref{phase1}, we present the full phase diagram.

A similar reasoning can be applied to other weight functions.
For instance, for $K(x)\propto x^{-\nu}$ (long-range interactions) and $p(m)=e^{U \delta_{m0}}$, the corresponding probabilities are
\ba
	\ln P_{\rm smooth} \sim -W - \nu W \ln (N/W^2), \\
	\ln P_{\rm rectangular} \sim -W - \nu \ln (N/W),
\ea
and the latter probability is always larger, provided that $W\approx {\rm const}$, so the shape turns out to be always rectangular. Similarly to finite-size scaling at first-order transitions, the height scales with $N$, while the width stays approximately constant, corresponding to a correlation length for first-order transitions that stays finite.
For another choice, $K(x)\propto x^{-\nu}$ and $p(m)\propto m^{-b}$, one can show that the condensate emerges only if $b>0$ and $\nu>1-b/2$, as follows from Eq.~(\ref{eqh1}), applied to the eigenvector $\phi_m\sim m^{-b/2-\nu}$.
If the condensate exists, it is always localized, because the ratio $P_1/P_2$, calculated as before, behaves as $\sim N^b$ and grows to infinity for $b>0$. Even if one performs the sum over the difference $\epsilon$ in occupation numbers, $P_1$ is still larger than $P_2$ by a factor of at least $N^{b/2}$.
Details are postponed to Ref.~\cite{bw1}.
Interestingly, from experimental observations of the shape and the scaling of the width with the system size, one can trace back the class of hopping interactions that are compatible with the observations.

To summarize, we have shown that a condensate which emerges in a PFSS as a result of spontaneous symmetry breaking, can be either extended or localized, depending on the competition between local and ultralocal interactions, and that its extension and the envelope can be calculated analytically. Let us close the paper with a remark on possible applications.
When atoms condense on a crystal surface, they can migrate and build extended islands. 
As experiments on the deposition of clusters \cite{exp} or fabrication of quantum dots \cite{qd} show, the islands can be extended in the direction perpendicular to the surface. 
It would be interesting to check if the PFSS model could be used to predict shapes and typical sizes of islands of atoms obtained in this way.
Also problems like mass transport on arbitrary networks could be addressed within the PFSS formalism. 

B.W. and W.J. thank the EC-RTN Network “ENRAGE” under grant No.~MRTN-CT-2004-005616 for support.
B.W. thanks the International Center for Transdisciplinary Studies (ICTS) at Jacobs University for support of some visits during this collaboration.

\end{document}